# SSGMT: A Secure Smart Grid Monitoring Technique

Sohini Roy

Arizona State University, Tempe AZ 85281, USA
`sroy39@asu.edu`

**Abstract.** Critical infrastructure systems like power grid require an improved critical information infrastructure (CII) that can not only help in monitoring of the critical entities but also take part in failure analysis and self-healing. Efficient designing of a CII is challenging as each kind of communication technology has its own advantages and disadvantages. Wired networks are highly scalable and secure, but they are neither cost effective nor dynamic in nature. Wireless communication technologies on the other hand are easy to deploy, low cost etc. but they are vulnerable to cyber-attacks. In order to optimize cost, power consumption, dynamic nature, accuracy and scalability a hybrid communication network is designed in this paper where a portion of the communication network is built using wireless sensor networks (WSN) and the rest is a wired network of fiber optic channels. To offer seamless operation of the hybrid communication network and provide security a Secure Smart Grid Monitoring Technique (SSGMT) is also proposed. The performance of the proposed hybrid CII for the generation and transmission system of power grid coupled with the SSGMT during different cyber-attacks is tested using NS2 simulator. The simulation results show that the SSGMT for a joint power communication network of IEEE 118-Bus system performs better than the prevailing wireless CIIs like Lo-ADI and Modified AODV.

## 1 Introduction

An improved and efficient Critical Information Infrastructure (CII) for a Critical Infrastructure System (CIS) gives birth to a smart CIS. It is the incorporation of features like full-duplex communication between the CII entities by the addition of embedded systems, automated metering in the smart homes, power distribution automation, pervasive monitoring etc., that converts a traditional power grid to a smart grid system. It is beyond any question that the CII of a smart grid must be accurate, scalable, and secure enough to instantly identify any kind of abnormal behavior in the entities of the power network, securely communicate that information to the control center and thereby help in taking necessary and timely action to ensure uninterrupted power supply.

As a result, finding the best suited design of a robust CII for smart grid has become a boiling topic of research. In [1] a crude idea of the design of a joint power-communication network is given using a test system consisting of 14 buses. Yet, the ground level details of the CII system are missing. The authors of [2], have come up with a realistic design of the CII of a smart grid by taking help from a power utility in the U.S. Southwest; their CII system relies completely on wired channels that either use SONET-over-Ethernet or Ethernet-over-Dense Wavelength Division Multiplexing. However, a completely wired CII is neither cost effective nor energy saving. Every CII entity in [2] draws power and thus a huge amount of power is devoted for monitoring the power network itself. Moreover, isolation of CII entities during a failure or a security threat and addition of a new entity in the network for hardening purpose or fault tolerance is extremely difficult and costly in a wired system.

Smart sensors like Phasor Measurement Units (PMUs) are already gaining popularity in smart grid system for measuring electrical waves. Power generation and transmission, power quality, equipment fitness, load capacity of equipment and load balancing in the grid can also be monitored by data sensing techniques. WSNs are comprised of low powered sensor nodes with easy installation process, lesser maintenance requirement, low installation cost, low power profile, high accuracy and scalability. All these have convinced the researchers that WSNs are a very good choice for the designing of the CII of a smart grid. However, the most common drawback of a sensor node is that it is battery powered and it is not easy to replace its dead battery. As a result, energy conservation becomes important. In the proposed work, energy efficiency is obtained by both energy aware routing technique for Supervisory Control and Data Acquisition (SCADA) data transmission and by the use of more expensive rechargeable Energy Harvesting Relay Nodes (EHRNs) for PMU data transmission to the Control Centers (CC). Also, the nodes and wireless channels are vulnerable to cyber-attacks. Some of the common cyber-attacks in WSNs are discussed in [3].

In this paper, a hybrid CII is designed in which a WSN based communication network is used between a sensing unit placed at a substation like a Remote Terminal Unit (RTU) or a PMU and a regional aggregation point like a Regional Sink node (RS) or Phasor Data Concentrator; and optical fiber based communication is used between the regional aggregation point and the CCs. SSGMT aims at securing the sensed data by means of light weight security protocols used in [4] like Elliptic-Curve-Public-Key Cryptography (ECC), Elliptic-Curve-Diffie-Helman Key exchange scheme (ECDH), Nested Hash Message Authentication Codes (NHMAC) and RC5 symmetric cypher.

The rest of the paper is structured as follows. Section 2 gives an overview of the CII system setup phase for SSGMT. Mitigation of different threats to the proposed network design is discussed in section 3 by adopting a secure routing technique. Section 4 does performance analysis of the proposed scheme SSGMT by means of comparing the simulation results with other existing secure remote monitoring technique for power grid like Lo-ADI [5] and Modified AODV [6]. Sections 5 concludes the paper and discusses the scope for future works.

## 2    Overview of the CII system setup phase for SSGMT

In order to provide a reliable remote monitoring technique for the smart grid, a generic hybrid CII system design is proposed in this section that can be applied on any given power network. In order to illustrate the steps of CII design, the generation and transmission part of a power grid formed by the IEEE 14-Bus system is considered.

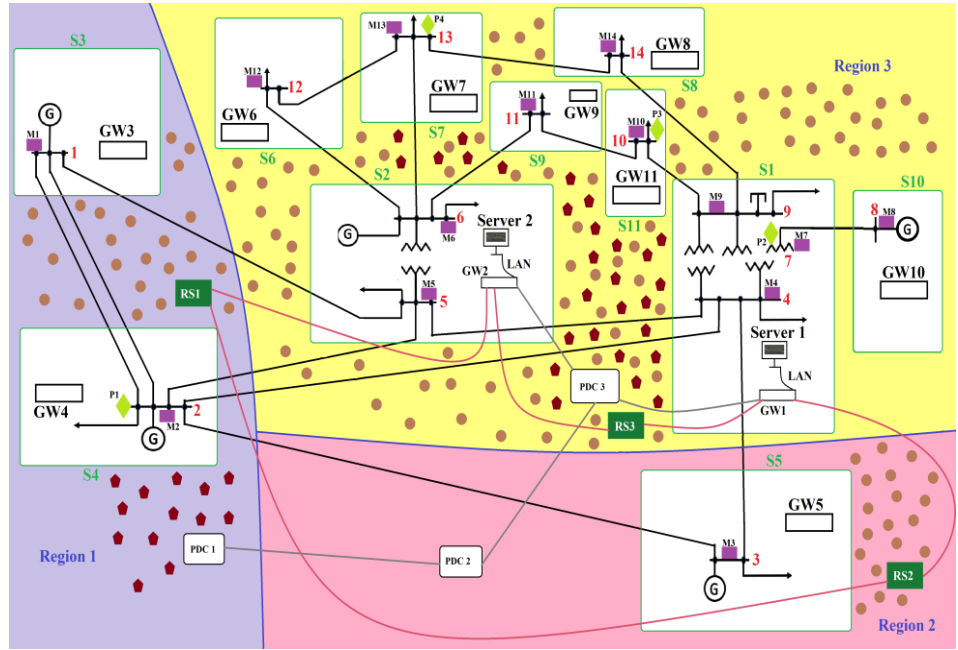

**Fig. 1.** Critical Information Infrastructure Design for a Smart Grid of IEEE 14-Bus

Initially, a given power grid is divided into several substations based on [7]. In Fig.1, the IEEE 14-Bus system is divided into 11 substations. After the substation division, two substations having the highest and second highest connectivity with other substations are selected as Main and backup CCs respectively. As the CCs are selected, all substations are equipped with a router acting as a gateway ($GW_i$) and a substation server is placed in each CC acting as the access point for the operator. The CC- gateways can receive optical signals from the optical channels and convert them to electrical signals using a photodiode and send those data to server via a wired LAN connection. Other $GW_i s$ can receive data from the sensors in that substation via Zigbee and forward that to either the non-rechargeable relay nodes or the EHRNs.

The distance between all pairs of substations ($S_i$ & $S_j$) is calculated as in [2]. Now, starting from a substation $S_i$ with the maximum connectivity among the border substations, all other substations which are within a given distance D of $S_i$, are marked as substations of a common monitoring region $R_x$. D is determined on the basis of network size and average distance between substations. Then the next substation which is the closest to $S_i$ but beyond the given distance and which is not yet placed in a monitoring region, is selected and the same process is repeated. This process is continued till every substation is placed within a particular $R_x$. In Fig.1, $S_1$ is selected as the main CC and $S_2$ is selected as the backup CC. Substations $S_3, S_4, S_5$ and $S_{10}$ are at the borders of the smart grid area. Among them, $S_4$ has the maximum connectivity, therefore the region division starts from it and finally, the smart grid network for IEEE 14-Bus system is divided into 3 monitoring regions.

In SSGMT, two different types of Zigbee enabled smart sensors are considered for monitoring the power network entities. The first type is the Measuring Unit (MU) based smart sensors [8] connected to an RTU for measuring SCADA input data and the second type is the Zigbee enabled PMU-based smart sensors or ZPMU [9].

In this step, MU-based sensors ($M_i$) are placed at every bus but PMU-based sensors ($P_i$) are placed at some of the buses using an optimal PMU placement algorithm [7]. If there are multiple $M_i$s in a substation then RTU receives the data from all such $M_i$s before forwarding them to the substation gateway ($GW_i$). Low-cost, non-rechargeable battery enabled relay nodes are randomly dispersed across the network area. These relay nodes can carry the SCADA data to a RS placed at every monitoring region. Each RSs is either connected to a neighboring RS or a CC-gateway via optical fiber channels to form a ring structure in order to provide fault tolerance. These RSs now convert the electrical signals obtained from the relay nodes to optical signals using a light emitting diode, associated with each RS. These optical signals are then carried to the CC-gateways via optical fiber channels and other RSs in the ring. TCP based communication is used between the RSs and the CC-gateway.

A phasor data concentrator (PDC), responsible for receiving and accumulating PMU data from multiple PMUs, is also placed at each region of the smart grid and few EHRNs are randomly deployed across the smart grid region. The idea behind the deployment of the two kinds of sensor nodes is that, the cheaper non-rechargeable relay nodes will follow an event-driven hierarchical routing approach to send the SCADA data and the EHRNs will always be active to accumulate synchrophasor data from the substations of each region and send the data to the local PDCs and finally to the CC-gateways. Due to the high volume of PMU data transfer from each substation having a PMU, the sensor nodes carrying them should always be active. IEEE C37.118 standard is maintained for communication of PMU data to the PDCs. PDCs can convert the data to optical signals in the similar way as RSs and send that to the CC-gateways either directly or via other PDCs in neighboring regions. PDCs also use TCP based communication to send the optical data to CCs.

## 3     Secure Smart Grid Monitoring Technique (SSGMT)

The goal of the CII for a smart grid is to securely transmit the sensed data from the sensors to the CCs and help in remote monitoring of the power grid. In order to achieve this with the help of a hybrid CII, the SSGMT is divided into 3 modules and described in this section.

## 3.1 Module 1: Data sensing by substation sensors and forwarding to substation gateways

In the first module of SSGMT, the $M_i$ and $P_i$ sensors placed in the substations sense electrical waves from the buses they are placed on and use Zigbee to send the data to the substation gateway $GW_i$. No security measure is adopted in this step as it is assumed that no cyber attack can harm the communication within a substation.

## 3.2 Module 2: Data forwarding by substation gateways to RSs and PDCs

The next phase of the hybrid CII system of SSGMT is data forwarding by $GW_i s$. $GW_i$ Use two separate methods for forwarding $M_i$ and $P_i$ data. First, the trust values ($TV_i$) of the non-rechargeable nodes ($N_i$) and EHRNs are determined by the $GW_i s$ of that region by means of forwarding a number of test messages through them to the RSs and PDCs of that region respectively. The $TV_i$ of each node is calculated using eq.1.

$$TV_i = \frac{MSG_{delivered}}{MSG_{sent}} * 100 \qquad (1)$$

It is assumed that all the CII entities are provided with a global key $GBK$ which an attacker cannot get hold of even if the entity is compromised. Also, a unique set of elliptic curves is stored in the memory of each CII entity for the purpose of ECC and ECDH protocols [4]. Also, in order to achieve those mechanisms, it is assumed that any pair of entities in the network agrees upon all the domain parameters of the elliptic curves stored in them. Now, rest of module 2 is described using the flowchart below.

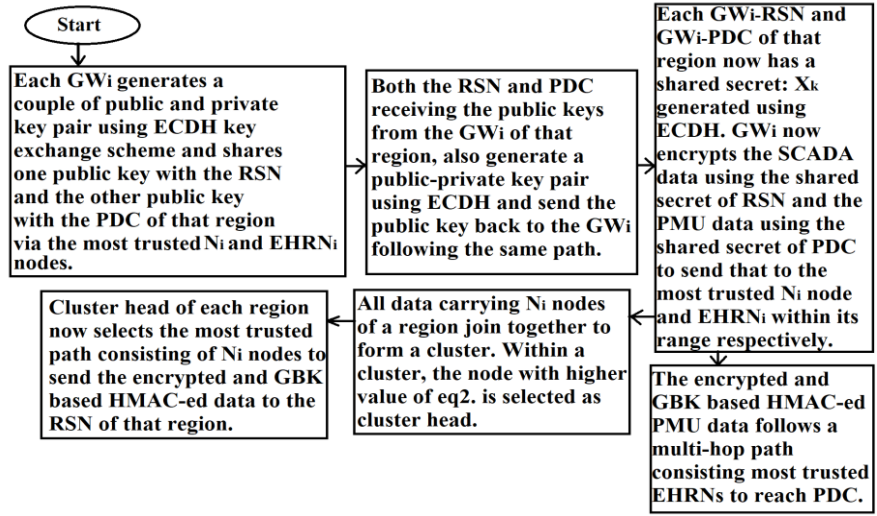

RC5 symmetric cipher [4] is used for the encryption of data using the shared secret. The GBK is used by $GW_i$ to generate a Hashed Message Authentication Code (HMAC) [4] over both the encrypted SCADA and PMU data and attached with the encrypted data for the purpose of authentication of data. Eq 2. is used to select the cluster head. $N_i$ nodes receiving data from $GW_i$ and with highest Candidate Value ($CV_i$) is selected as the cluster head.

$$CV_i = BP_i * TV_i * Cn_i \qquad (2)$$

In eq.2, $BP_i$ represents remaining battery power of $N_i$, $TV_i$ is the current trust value of $N_i$ and $Cn_i$ is the connectivity of $N_i$ with other nodes in the region. The same path to send data to RSs and PDCs is followed by each $GW_i$ until the RSs or PDCs send a rerouting request to the corresponding $GW_i$.

### 3.3 Module 3: Data forwarding by RSs and PDCs to CC-gateways

In this module, the RSs and PDCs after obtaining the encrypted and HMAC-ed data from the $Ns$ and EHRNs use the shared secret obtained for that sender $GW_i$ to decrypt the data packets. They also match the HMAC attached with the encrypted data to check if any false data injection took place. In case, the HMAC does not match, the data packet is dropped, and rerouting request is sent back to the sender. The main CC-server use ECC based public key cryptography [4] and generate a public key for encryption and a private key for decryption of data. The ECC based public key of the main CC-server is sent to each of the RSs through the RS-ring and also to the PDCs via other PDCs and the optical channels. The main CC-gateway use a dedicated and secure optical channel to communicate with the backup CC-gateway. This channel is used to share the private key with the backup CC-server. RSs are responsible for data aggregation. Aggregated SCADA data from the $Ns$ are encrypted by the RSs using the public key of the main CC-server. This encrypted data is sent to both the CC-gateways via the RS-ring. In the similar way PDCs send the aggregated and encrypted synchrophasor data via other PDCs to the CC-gateways wherefrom they reach the CC-servers.

## 4 Simulation Results

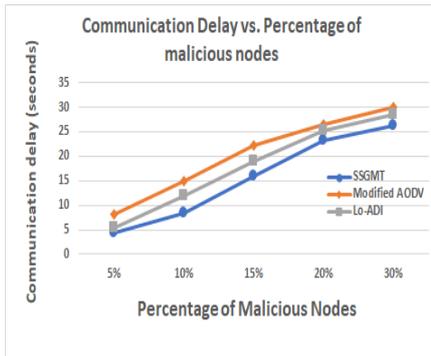 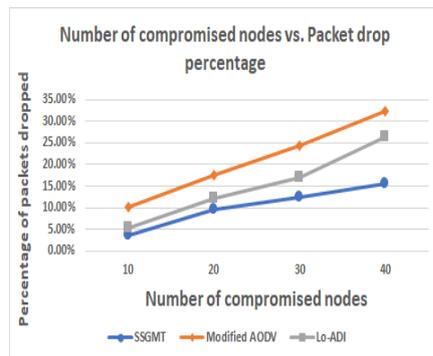

**Fig. 2.** Communication delay vs. Malicious nodes     **Fig. 3.** Number of compromised nodes vs. Packet drop

In this section, the CII for a smart grid network of IEEE 118-Bus system is considered. The total network region is divided into 8 regions and the power grid is divided into 107 substations. Substation 61 is selected as the main CC and it consists of 3 buses─68,69 and 116. Substation 16, consisting of buses─17 and 30, is selected as the backup CC. In order to analyze the performance of SSGMT in this network setup, a total of 1500 non-rechargeable relay nodes, 500 EHRNs and 8 PDCs are deployed in the network area and NS2.29 is used for simulation. The simulation results are compared with existing WSN based CII systems like Lo-ADI [5] and modified AODV [6].

## 5    Conclusion and Future Works

The region based remote monitoring adopted by SSGMT helps in easy identification of a failure in the power grid or an attack in the communication network of the smart grid. SSGMT obtains data privacy by the encryption/decryption mechanisms, data integrity and authenticity by the HMACs. Delay, security, power consumption, scalability etc. are optimized by this hybrid network design of SSGMT. Designing a threat model with attacks on smart sensors, gateways or servers and analyzing the effect of cyber-attacks on power grid can be another direction of future work.